\documentclass[aps,pre,preprint,showpacs]{revtex4-1}
\usepackage{xcolor}
\usepackage[hypertex]{hyperref}
\usepackage{graphicx,subfigure}
\usepackage{amsmath,amsfonts,amssymb}

\newcommand{\f}{\frac}
\newcommand{\suml}{\sum\limits}

\newcommand{\intl}{\int\limits}

\graphicspath{{D:/localtexmf/Mytex/LGH/mathplots/}}
\begin{document}
\title{Generalized principle of
corresponding states and the scale invariant mean-field approach}
\author{L.~A. Bulavin}
\email{bulavin@univ.kiev.ua}
\affiliation{Department of Molecular Physics, \\
Taras Shevchenko National University of Kyiv, 2,
Prosp. Academician Glushkov, Kyiv 03022, Ukraine}
\author{V.L. Kulinskii}
\email{kulinskij@onu.edu.ua}
\affiliation{Department for Theoretical
Physics, Odessa National University, Dvoryanskaya 2, 65026 Odessa, Ukraine}
\begin{abstract}
In this paper we apply the global isomorphism approach [V.~L. Kulinskii, J. Phys. Chem. B \textbf{114} 2852 (2010)] between the Lennard-Jones fluids and Lattice Gas model
to the study of the liquid-vapor equilibrium for the systems with the short-ranged potentials like Buckingham and the  $Mie$-potentials. The estimates for the critical point locus correlate quite well with the available numerical data. Also within the proposed approach we give the explanation for the correlation between the value of the second virial coefficient at the critical temperature and the particle volume found in [G. A. Vliegenthart and H. N. W. Lekkerkerker, J. Chem. Phys. \textbf{112} 5364 (2000)].
\end{abstract}
 \maketitle
\section{Introduction}\label{sec_intro}
The Principle of the Corresponding States (PCS) is one of the driving force in searching the unifying description in the variety of the thermodynamic properties of the complex matter. From the microscopic point of view the PCS is based on simple scaling properties of the interaction potential which is usually assumed to be of pairwise character. Actually the microscopic interactions in real substances are more complex and do not conform with the conditions at which the PCS can be derived rigorously from the first principles \cite{book_prigozhisolut,eos_pcspitzer_jcp1939}. The simple scaling form of the PCS was extended to include more parameters which are connected with the most important properties of the interparticle interactions. For example K.S. Pitzer \cite{liq_pcspitzer1_jamchemsoc1955} generalized the PCS by inclusion the acentric factor $\omega$ to account for the nonsphericity effects in the interaction. In connection with this a lot of model potentials have been invented for the description of the thermodynamic behavior of the fluid systems. Owing to the fast computers and sophisticated numerical algorithms now one can relatively easily to get the thermodynamic properties of complex matter from simulations \cite{book_frenkelsimul}. The detailed description of the bulk liquid-vapor coexistence as well as interface is possible for the model potentials in addition to usual structural measurements in real liquids. The most crucial property with respect to which the qualitative division of the potentials into distinct classes of the thermodynamic behavior has the physical meaning is the effective range of interaction. Indeed, it is known from the renormalization group treatment \cite{crit_fisherlongshort_prl1972} the interactions decaying faster than $r^{-(d+s)}\,,s>2-\eta_{Is}$, where $\eta_{Is}\approx 0.033$ is the Fisher's critical exponent lead to the Ising-like critical behavior, where the fluctuations play essential role.

Recently in series of works of E.~Apfelbaum and V.~Vorob'ev \cite{eos_zenoapfelbaum_jchempb2006,*eos_zenoapfelbaum_jchempb2008,
eos_zenoapfelbaum1_jpcb2009} new correlations between the locus of the critical point (CP) and the line defined by the equation
\begin{equation}\label{z1}
Z = \f{P}{n\,T} = 1\,,
\end{equation}
where $P,T,n$ are the pressure, the temperature and the density correspondingly, have been discovered. This line is called the Zeno-line. Along this line the compressibility factor $Z$ is equal to unit which means that the ideal gas equation of state $P=n\,T$ is fulfilled. For the classical van der Waals (vsW) equation of state (EoS) \cite{book_ll5_en}:
\[P = \f{n\,T}{1-n\,b}-a\,n^2\,,\]
the relation \eqref{z1} leads to the perfectly straight line:
\begin{equation}\label{z1vdw}
  \f{n}{n_b}+\f{T}{T^{(vdW)}_{B}} = 1\,,
\end{equation}
where $n_b = 1/b$ and $T^{(vdW)}_{B} = a/b$ is the Boyle temperature in the van der Waals approximation \cite{book_ll5_en}. The relation \eqref{z1vdw} is known as the Batchinsky law \cite{eos_zenobatschinski_annphys1906} (see also \cite{eos_zenobenamotz_isrchemphysj1990}). As has been shown by Apfelbaum and Vorob'ev this law is fulfilled with quite good accuracy for wide variety of substances as well as the model systems \cite{eos_zenoline_potentials_jcp2009,eos_zenoapfelbaum1_jpcb2009}. The authors themselves noted that their results can be thought in terms of some extension of the PCS because they can be \textit{applied to a wider group of substances in comparison with ones satisfying
the corresponding states principle} \cite{eos_zenoapfelbaum1_jpcb2009}. Of course it is the symmetry which underlies PCS itself as well as any of its possible extension. The classical PCS is based on the simple scaling symmetry of the Hamiltonian. But the empirical facts of \cite{eos_zenoapfelbaum_jchempb2006,eos_zenoapfelbaum_jchempb2008,eos_zenobenamotz_isrchemphysj1990} testify that the similarity of the thermodynamic properties has much wider applicability then one can get from the trivial rescaling of the parameters of interaction.

The straightness of the Zeno-line \eqref{z1} seems rather special although the deviations are small at least for the simple molecular liquids and for model potentials \cite{eos_zenoapfelbaum_jchempa2008,eos_zeno_jphyschemb2000}.
As was shown in \cite{eos_pcsongmason_jcp1991,*eos_zenoline_jphyschem1992} \eqref{z1vdw} is equivalent to the linear density dependence of the inverse value of the binary correlation function at contact.
It is easy to see that in low density region the line \eqref{z1} indeed is close to the straight line. The latter intersects the $T$-axes at the Boyle point $T_B$ determined by the condition:
\begin{equation}\label{b2}
B_2(T_B)= 0\,,
\end{equation}
where $B_2$ is the second virial coefficient \cite{book_hansenmcdonald}.

In addition to \eqref{b2} the slope of the line should be determined so that it will be the tangent to the extrapolation of the binodal. Usually, this is fixed by the condition \cite{eos_zeno_jphyschemb2000}:
\begin{equation}\label{TZ}
n_b= T_B  \,\f{B'_2\left(\,T_B\,\right)}
{B_3\left(\,T_B\,\right)}\,.
\end{equation}

The search for the symmetry which generalizes the classical PCS and underlies the empirical findings in \cite{eos_zenoapfelbaum_jchempb2006,eos_zenoapfelbaum_jchempb2008,eos_zenobenamotz_isrchemphysj1990}
can be done within the approach which was proposed in \cite{eos_zenome_jphyschemb2010} and further extended in \cite{eos_zenomeglobal_jcp2010}.
It is based on the global isomorphism between liquid-vapor part of the phase diagram of the molecular fluid and the Lattice Gas (LG) or the Ising model. It is defined by the Hamiltonian
\begin{equation}\label{ham_latticegas}
  H = -J\suml_{
\left\langle\, ij \,\right\rangle
  } \, n_{i}\,n_{j} - \mu \,\suml_{i}\,n_{i}\,.
\end{equation}
Here $n_{i} = 0,1$ whether the site is empty or occupied correspondingly. The quantity $J$ is the energy of the site-site interaction of the nearest sites $i$ and $j$, $\mu$ is the chemical potential. We denote by $t$ the temperature variable corresponding to the Hamiltonian \eqref{ham_latticegas}. The order parameter is the probability of occupation of the lattice site $x = \left\langle\, n_i
\,\right\rangle$ and serves as the analog of the density.

The aim of this work is to apply the above mentioned approach to the prediction of the locus of the CP for the widely used potentials and to compare the results with known \cite{eos_zenoapfelbaum1_jpcb2009}.

\section{The generalized principle of corresponding states}
  In \cite{eos_zenome_jphyschemb2010} it was shown that the transformation of to the LG-variables $(x,t)$ of the form:
\begin{equation}\label{projtransfr_nx}
  n =\, n_b\,\f{x}{1+z \,t}\,,\quad
  T =\, T_Z\,\f{z\, t}{1+z \,t}\,,
\end{equation}
allows to map the liquid-vapor part of the phase diagram for the molecular system to the phase diagram of the lattice model. In particular the \eqref{projtransfr_nx} maps the line $x=1$ which represents the ``holeless`` state of the LG onto the tangent line to the liquid branch of the binodal at $T\to 0$ \cite{eos_zenome_jphyschemb2010}. In fact the transformation \eqref{projtransfr_nx} was proposed in \cite{eos_zenome_jphyschemb2010} as the geometrical reformulation of these empirical facts.

Here the parameter $z$ can be defined by the correspondence between the loci of critical points of fluid $(n_c,T_c)$ and the lattice gas $x_c=1/2\,, t_c=1$:
\begin{equation}\label{z_c}
  z = \f{T_c}{T_Z - T_c}\,.
\end{equation}
The choice of the parameters $T_Z$ and $n_b$ will be discussed below in Section~\ref{sec_cplocus}.

The coordinates of the CP for the liquid are:
\begin{equation}\label{cp_fluid}
  n_{c} = \f{n_b}{2\left(\,1+z\,\right)}\,,\quad   T_{c} = T_Z\, \f{z}{1+z}\,.
\end{equation}
If $T_Z$ and $n_b$ are fixed then $z$ parameterizes the locus of the CP:
\begin{equation}\label{cpline}
  \f{n_c}{n_b} +   \f{T_c}{T_Z} = \f{z +1/2}{1 +z}
\end{equation}
This means that for the substances belonging to the same class of the corresponding states represented by $z$,  the loci of the critical points  scaled to the $T_Z$ and $n_b$ lie along the straight line. This correlates with the empirical analysis in \cite{eos_zenoapfelbaum_jchempb2008}. Note that the specific form of the transformation \eqref{projtransfr_nx} and all results derived from it rely solely on the geometrical facts about phase diagrams of the Lattice Gas and the fluid. If the global isomorphism is accepted, \eqref{cpline} can be considered as  the generalized PCS for the systems isomorphous to the Lattice Gas. The parameter $z$ parameterizes  the classes of corresponding states. The latter are defined in common way via the scaling of the parameter of the interaction \cite{eos_zenomeglobal_jcp2010}.

Of course \eqref{projtransfr_nx} is not the exact transformation, nevertheless, as it is shown in \cite{eos_zenomeglobal_jcp2010}, this gives the consistent framework for the description of the known empirical facts and allows to get rather good estimates for the locus of the CP directly from the interparticle potential. It is based on the simple scaling relations
$
n_c\sim \f{1}{r^d_c}\,, T_c \sim |\Phi_{att}(r_c)|$, where $r_c$ is the mean interparticle distance in the critical state. The relation
\begin{equation}\label{cp_scaling_z}
- \f{d\,\ln{\left(\,T_c/T_Z\,\right)}}{d\,\ln{\left(\,n_c/n_b\,\right)}} = \f{1}{z}\,,
\end{equation}
follows from \eqref{cp_fluid} and leads to:
\begin{equation}\label{cp_locus_potential}
\f{d\,\ln \left(\,\left|\Phi_{att}(n^{-1/d}_c)\right|/T_Z\,\right)}{d\,\ln \left(\,n_c/n_b\,\right)} = \f{1}{z}
\end{equation}
If the attractive part of the potential has the power-like asymptotic $\Phi_{att}(r)\propto r^{-(d+s)} $ then  $1/z =1+s/d$. E.g. for three dimensional LJ potential $z = 1/2$.
The key point is the determination of the Zeno-line parameters $n_b$ and $T_Z$. Below we describe our approach to the determination of these parameters.

It is still seems quite unusual that the coordinates of the the critical point which belong to the dense fluid region are correlated with the properties derived for low density states \cite{eos_zenobenamotz_isrchemphysj1990}. We believe that this fact is reflected by the global character of the proposed transformation \eqref{projtransfr_nx}. In the end of this section we show how the proposed approach allows to clarify the essence of another example of correlation between critical parameters and the low density properties which was discovered in \cite{eos_seconvircrit_jcp2000}. It was demonstrated that the critical temperature for many three dimensional LJ-like systems obey the relation:
\begin{equation}\label{b2virial6lekker}
  B_2(T_c) \approx -C\,v_0\,,
\end{equation}
with $C\approx 6$, $v_0 = \f{\pi}{6}\,\sigma^3$ and $\sigma$ is the distance where the potential $\Phi(r)$ crosses zero $\Phi(\sigma) = 0$. Here
\begin{equation}\label{b2virial}
    B_2(T) = 2\,\pi\,\intl_{0}^{\infty}\,\left(\,1-e^{-\Phi(r)/T}\,\right)\,r^2\,dr
\end{equation}
is the second virial coefficient. We call \eqref{b2virial6lekker} the Vliegenthart-Lekkerkerker relation. The results of \cite{eos_seconvircrit_jcp2000} allow to say that the value of $C$ for the LJ potentials appears to be close to $2\pi$. Further we will use this value.

It seems plausible that the analog of \eqref{b2virial6lekker} should exist for other model systems. In such case the parameter $C$ represents the class of corresponding states. Unfortunately, the exact connection of the value $C$ with the specific properties of the interaction potential remains unclear. To clarify this issue we note that in accordance with \eqref{cp_fluid} for the LJ fluids the approximation
\begin{equation}\label{tctvdw}
 T_c\approx T^{(vdW)}_B/3\,.
\end{equation}
can be used. Taking the formal high-temperature expansion of \eqref{b2virial} at $T\to \infty$ (see e.g. \cite{book_ll5_en}) we get:
\begin{equation}\label{b2virial_vdw}
    B_2(T_c) = b \left(1-T^{(vdW)}_B/T_c\right)+ \ldots
\end{equation}
where all higher order terms are omitted. Since $b = 4v_0$ we get the estimate:
\begin{equation}\label{b2_me}
    B_2(T_c) = -8\,v_0+\ldots\,,
\end{equation}
which is quite satisfactory explains the value of factor $C$ in Vliegenthart-Lekkerkerker relation \eqref{b2virial6lekker}. It can be shown using the specific form of the potentials \eqref{miemn} that the higher order terms increase the zeroth order coefficient in \eqref{b2_me}.

\section{The locus of the critical point}\label{sec_cplocus}
One of the most fascinating guesses of Apfelbaum and Vorob'ev
was that the Zeno-line should be the tangent to the extrapolation of the binodal into the unphysical region $T\to 0$ where the liquid phase ceases to exist as the stable state \cite{eos_zenoapfelbaum_jchempb2008}. This assumption allowed to consider much wider class of substances than in classical PCS \cite{eos_zenoapfelbaum1_jpcb2009} and formulate the conception of the Triangle of Liquid-Gas states \cite{eos_zenoapfelbaum_jchempb2006}.

It is rather tricky to extrapolate the liquid branch of the binodal into the region of  solid state. This is the region which determines the parameter $n_b$. The natural parameter which restricts the applicability of this procedure is the locus of the triple point.

The isomorphism transformation \eqref{projtransfr_nx} relies heavily on the correspondence between the tangent to the liquid branch of the binodal and the line $x=1$ which represents the ``holeless`` state of the LG \cite{eos_zenome_jphyschemb2010}. Since any EoS can be approximated by the vdW EoS for which the Zeno-line is exactly the straight line we use the the generic vdW approach to the EoS proposed in \cite{eos_genvdw_pre2001,*eos_genvdw1_jcp2001} for the short ranged potentials. It is based on the fact that the EoS for molecular fluids can be written as following:
\begin{equation}\label{p_basic}
  p = p_{+}(n,T)+p_{-}(n,T)\,,
\end{equation}
where $p_{+}$ is the pressure contribution due
to hard core repulsive interactions and $p_{-}$ is the
contribution of the attractive long range part of the
potential. These two contributions are represented in
the van der Waals (vdW) form:
\begin{equation}\label{p_repatt_vdw}
  p_{+} = \f{n\,T}{1- B\,n}\,,\quad p_{-} = -A\,n^2\,.
\end{equation}
E.g. for the potentials with the hard core the coefficients $A$ and $B$ are given by
\cite{eos_genvdw_pre2001}:
\begin{align}
A=& -\f{2\pi \sigma ^3}{3}\,T\,\intl_{\sigma}^{+\infty} dr\,r^3\,y(r,n,T)\,f'(r,n,T)\,,\label{a_genvdw}\\
  B =& \,b\,\f{y(\sigma ,n,T)}{1+b\,n\,y(\sigma ,n,T)}\,, \label{b_genvdw}
\end{align}
where $\sigma$ is the diameter of the hard core, $f = e^{-\beta v(r)}-1$ is the Mayer function, $y(r,n,T) = e^{\beta v(r)}\,g(r,n,T)$ - the cavity function \cite{book_hansenmcdonald}, $g(r,n,T)$ - pair distribution function. Thus the vdW-like coefficients $A$ and $B$ are in general the functions of the thermodynamic state which are in general nonanalytic \cite{eos_genvdw1_jcp2001}. Once the cavity function is determined the vdW-like coefficients can be obtained as the corresponding low density limiting values:
\begin{align}\label{vdw_ab}
  \lim_{n\to 0} A =& \,\,a + O(1/T)\,,\quad b = \lim_{n\to 0}\, B \,,\\
a =&\,\, \pi\,\intl_{\sigma}^{+\infty}\Phi(r)\,r^2\,dr
\end{align}
because if $n\to 0$ then $y(r,n,T) \to 1$. Thus
the Boyle temperature in the vdW approximation is:
\begin{equation}\label{tbvdw}
  T^{(vdW)}_B  = \f{a}{b}\,.
\end{equation}


The parameter $n_b$ is determined by the equation analogous to the one used for the Zeno-line \cite{eos_zenobenamotz_isrchemphysj1990} with the change $T_B\to T_Z = T^{(vdW)}_{B}$:
\begin{equation}\label{nbtb}
n_b= T_Z  \,\f{B'_2\left(\,T_Z\,\right)}
{B_3\left(\,T_Z\,\right)}\,.
\end{equation}

In this section we apply the mean-field scaling approach proposed in \cite{eos_zenomeglobal_jcp2010} for the determination of the CP locus. It is based on the scaling properties of the attractive part of the potential and the global isomorphism and gives rather good estimates for the LJ fluids.

We consider the set of generalized LJ potentials (so called $Mie(m,n)$ potentials) \cite{eos_miepotential_annphysik1903} (see also \cite{eos_ljmie_intjthermophys1980}):
\begin{equation}\label{miemn}
  \Phi(r;m,n) = \varepsilon\f{m}{n-m}\,\left(\, \f{n}{m}\,\right)^{n/(n-m)}\,\left(\,\left(\,\f{\sigma}{r}\,\right)^n - \left(\,\f{\sigma}{r}\,\right)^{m}\,\right)\,,
\end{equation}
Note that at $n\to \infty$ we get the Sutherland potential
\begin{equation}\label{sutherland_potetial}
  \Phi_{S}(r;m) =
  \begin{cases}
    \infty & \quad r\le \sigma , \\
    -\f{\varepsilon}{r^{m}} & \quad r>\sigma\,\, .
  \end{cases}
\end{equation}
Further we pay special attention to the case $m=6$, for which the extensive numerical results are available \cite{crit_liqvamiepotent_jcp2000,crit_supercrit_physb2001,crit_ljmn_physleta2008}.
In particular, the results obtained for the liquid-vapor coexistence curve in \cite{crit_liqvamiepotent_jcp2000} for the set $\Phi(r;6,n)$ with $7\le n\le 32$ testify that scaled to the critical parameters these curves practically coincide.
Thus the PCS holds in this case. This fact stipulates the search of the scale symmetry which connects the critical temperatures for $\Phi(r;6,n)$ with different $n$.

This can be done on the basis of the scaling property of the attractive part of the potential since both $T_Z$ and $T_c$ are determined by it in accordance with \cite{eos_zenome_jphyschemb2010}. Since $T_Z$ is determined by the $T^{(vdW)}_B$ there is the arbitrariness in the choice of the effective diameter of the particle. This scale represents the separation of the regions attributed to the repulsive and the attractive part of the potentials. Usually, it is determined by the distance where the potential crosses zero. For the potentials \eqref{miemn} it is given by $\sigma$. Then
the potentials $\Phi(r;6,n)$ differ only by the scale factor at $r^{-6}$ and in accordance with \eqref{tbvdw} we obtain:
\begin{equation}\label{tbmnsc}
  \f{T^{(vdW)}_{B}(6,n_1)}{T^{(vdW)}_{B}(6,n_2)} = \sqrt{\f{n_2-6}{n_1-6}}\, \left(\,\f{n_1}{6} \,\right)^{\f{n_1}{2(n_1 - 6)}}\left(\,\f{n_2}{6} \,\right)^{-\f{n_2}{2(n_2 - 6)}}
\end{equation}
We choose the standard LJ potential $\Phi(r;6,12)$ as the reference case.
Then from simple rescaling we get the dependence of the critical temperatures for $\Phi(r;6,n)$:
\begin{equation}\label{tcscale_lj}
  T_{Z}(n) =2\,\sqrt{\f{6}{n-6}}\,\left(\,\f{n}{6}\,\right)^{\f{n}{2(n-6)}}\quad \Rightarrow\quad  T_{c}(n) =\f{2}{3}\sqrt{\f{6}{n-6}}\,\left(\,\f{n}{6}\,\right)^{\f{n}{2(n-6)}}
\end{equation}
It is interesting to note that both estimates \eqref{b2virial6lekker} and \eqref{tctvdw} practically coincide for the standard LJ potential $\Phi(r;6,12)$. The relation
\begin{equation}\label{me_vl_cond}
  T^{vdW}_{B}(6,n) = 3\,T^{(VL)}_{c}(n)\,,
\end{equation}
gives $n \approx 11.94$, which is very close to $n=12$.
Note that Vliegenthart-Lekkerkerker relation
\eqref{b2virial6lekker} does not depend on the choice
of the distance separating the attractive and repulsive
regions, i.e. it is scale invariant.

\begin{figure}
\center
\includegraphics[scale=1]{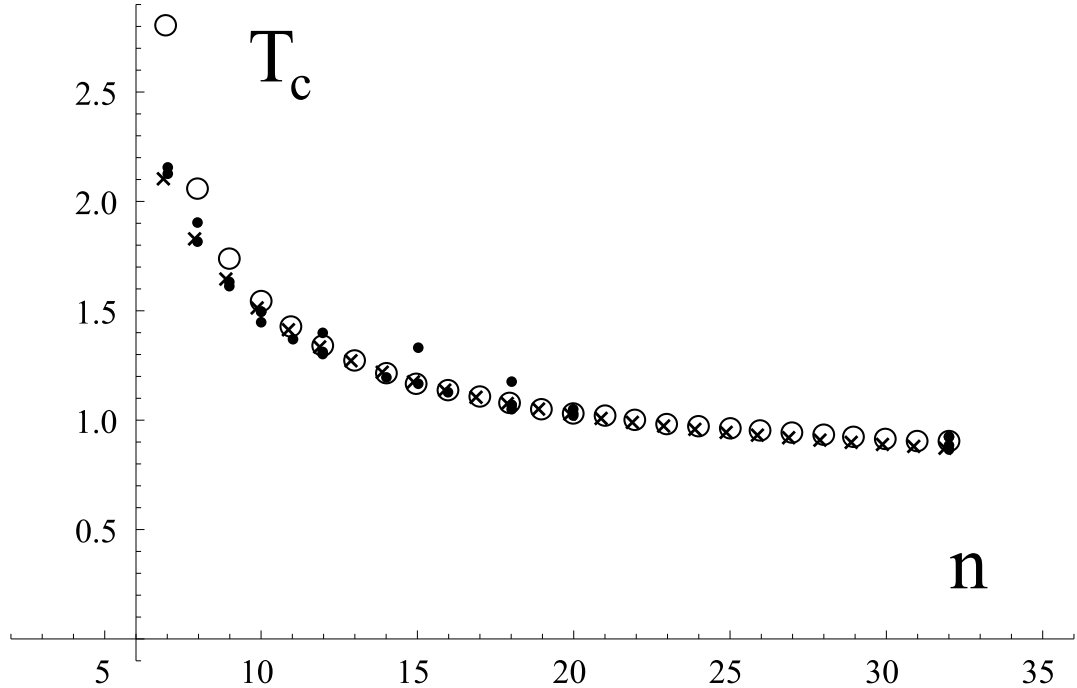}
  \caption{Critical temperature $T_c$ of $Mie(6,n)$-fluid: theory (open) and the numerical data (filled) \cite{crit_liqvamiepotent_jcp2000,*eos_latticecontinuum_jcp2000,*liq_ljphasediagr_jcp2005}. Also the results for $T_c$ obtained with the help of  the Vliegenthart-Lekkerkerker relation \eqref{b2virial6lekker} with $C = 2\pi$ are shown (crosses).} \label{fig_tc}
\end{figure}

\begin{figure}
\center
\includegraphics[scale=1]{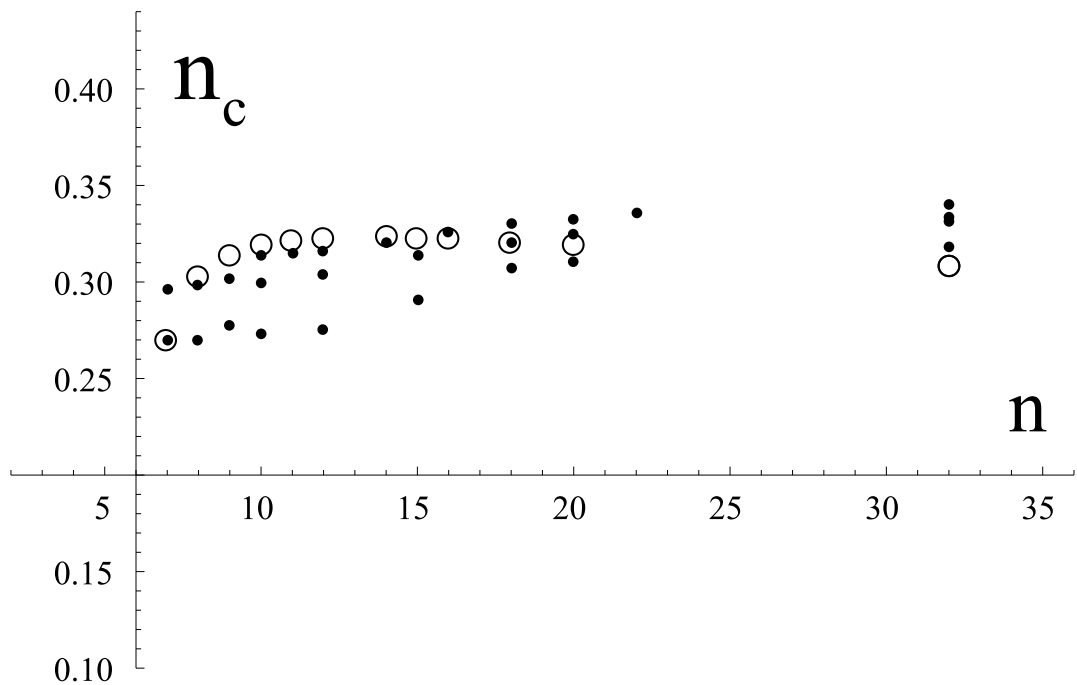}
  \caption{Critical density $n_c$ of $Mie(6,n)$-fluid: theory (open circles) and the numerical data (points) \cite{crit_liqvamiepotent_jcp2000,*eos_latticecontinuum_jcp2000,*liq_ljphasediagr_jcp2005}.}\label{fig_nc}
\end{figure}

The values for the critical density $n_c(n)$ are calculated according to \eqref{nbtb}.
The results are shown in Table~\ref{tab_lj} and
Figs.~\ref{fig_tc},\ref{fig_nc}. For completeness we
give the values of $T_c$ obtained using the
Vliegenthart-Lekkerkerker relation \eqref{b2virial6lekker}.

\begin{table}\label{tab_lj}
\center
\begin{tabular}{|c|c|c|c|}
  \hline
  $n$ & $n_c$ & $T_c$ &$T^{(VL)}_c$\\
\hline
7&0.269&2.801&2.107\\
\hline
8&0.303&2.053&1.832\\
\hline
9&0.314&1.732&1.648\\
\hline
10&0.319&1.546&1.516\\
\hline
11&0.321&1.423&1.417\\
\hline
12&0.322&1.333& 1.338\\
\hline
14&0.323&1.212& 1.223\\
\hline
15&0.322&1.168& 1.178\\
\hline
16&0.322&1.132&1.141\\
\hline
18&0.32&1.075&1.079\\
\hline
20&0.319&1.031&1.032\\
\hline
32&0.308&0.897&0.875\\
\hline
\end{tabular}
\caption{The results for the locus of the CP obtained from the scale invariant mean-field approach. The values $T^{(VL)}_c$ obtained using Vliegenthart-Lekkerkerker relation \eqref{b2virial6lekker} with $C=2\pi$.}
\end{table}
The results for the Sutherland potential can be obtained from \eqref{tcscale_lj} by taking the limit $n\to \infty$. Also we put $n_b\sigma^3 = 1$ in view of natural hard core cut-off for this potential.
\begin{table}\label{tab_suth}
\center
\begin{tabular}{|c|c|c|c|c|c|}
  \hline
  $m$ & $n_c$ & $n^{(num)}_c$ & $T_c$ & $T^{(VL)}_c$ & $T^{(num)}_c$ \\
\hline
\hline
  3.1 & 0.254 & 0.247 & 14.75& 8.14 & 11.45 \\
\hline
  4 & 0.286 & 0.299 & 1.286 & 1.307 &1.37 \\
\hline
  6 & 0.333 & 0.376 & 0.667 & 0.574 &0.597 \\
  \hline
\end{tabular}
\caption{The results for the Sutherland potential ($Mie(6,n), \,n=\infty$) \eqref{sutherland_potetial}.}
\end{table}

The same procedure can be applied to the Buckingham potential:
\begin{equation}\label{buck_potetial}
  \Phi_{B}(r;a) =
  \begin{cases}
    \infty\,\,, & \quad x\le r_0/r_m , \\
   \f{\varepsilon}{1-6/a}\left(\f{6}{a}\,\exp\left(\,a\left(\, 1-x\,\right) \,\right) -\f{1}{x^{6}}\right)\,\,, & \quad x>r_0/r_m\,\, ,
  \end{cases}
\end{equation}
where $r_0$ determines the hardcore cut-off and is defined
by the point of maximum of $\Phi_{B}$. The distance $r_m$ is the point where $\Phi_{B}$ reaches minimum $\Phi_{B}(r_0) = -\varepsilon$. Also the distance where $\Phi_{B}$ crosses zero we denote $\sigma $. It depends on the parameter $a$ and clearly $r_0<\sigma < r_m$.
Comparing \eqref{miemn} for m=6 and \eqref{buck_potetial} we see that the corresponding change of scale is needed.
It is easy to find that at $a \approx 12.8$ the attractive part of the Buckingham potential equals to that of the standard LJ 6-12 potential $\Phi(r;6,12)$ and therefore for such potential $T_c  = T^{(vdW)}_{B}/3$. So we choose this case as the reference point. The analog of \eqref{tcscale_lj} is:
\begin{equation}\label{tcscale_buck}
  T_{c}(a) = \f{2}{3}\,\left(\, \f{r_m}{\sigma}\,\right)^3\,\f{1}{\sqrt{1 - 6/a}}
\end{equation}
The density made dimensionless in units of the distance where the potential has minimum $r_{m}$ \cite{eos_zenoapfelbaum1_jpcb2009}.

\begin{figure}
\center
\includegraphics[scale=1]{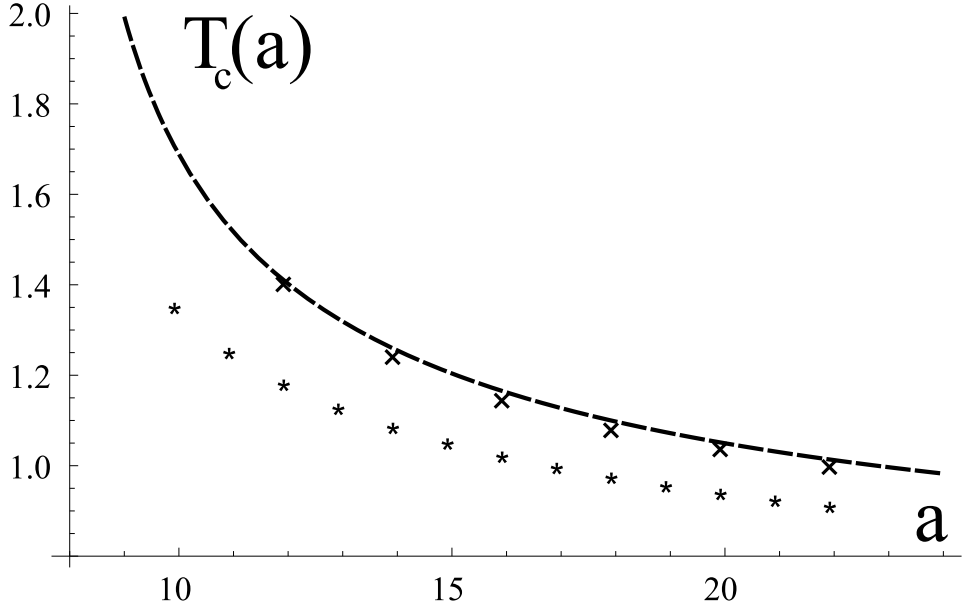}
  \caption{Critical temperature $T_c$ of Buckingham potential: theory (dashed) and the numerical data (crosses) \cite{eos_buckingham_jcp1998}, the Vliegenthart-Lekkerkerker relation \eqref{b2virial6lekker} with $C=2\pi$ (stars).} \label{fig_tc_buck}
\end{figure}
\begin{figure}
\center
\includegraphics[scale=1]{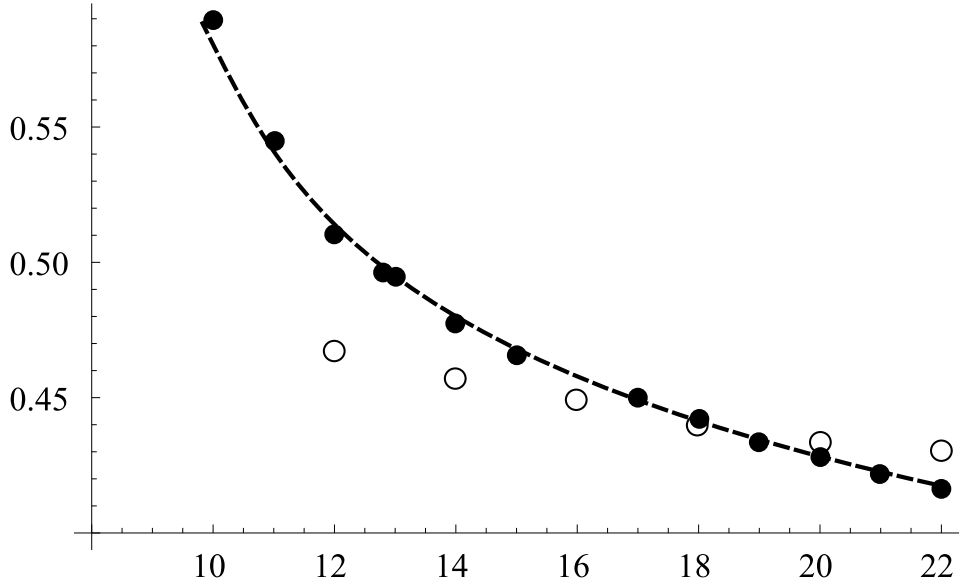}
  \caption{Critical density $n_c$ of Buckingham potential: theory (filled) and the numerical data (open) \cite{eos_buckingham_jcp1998}. The dashed line is the interpolation of the calculated values of $n_c$.} \label{fig_nc_buck}
\end{figure}
The comparison of the results and available numerical data for the Buckingham potential is given in Figs.~\ref{fig_tc_buck},\ref{fig_nc_buck}.

\section{Conclusions}
We considered the application of the scale invariant mean-field approach as the consequence of the global isomorphism between LG and LJ-like fluids to the calculation of the CP locus for such fluids. Simple scaling expressions for the critical temperatures are obtained for $Mie(6,n)$ and Buckingham potentials.
The comparison with the available numerical data shows satisfactory agreement. Also we proposed the explanation for the result of \cite{eos_seconvircrit_jcp2000}, namely, the constancy of the value of the second virial coefficient at the critical temperature. The presented results are based on the $\sim r^{-6}$ asymptotical behavior of the attractive part of the potential. For completeness it would be interesting to apply the proposed approach to the non power-like potentials like Yukawa or step-like potentials, e.g. square-well potential. We leave this for subsequent investigation.

In previous consideration we assumed that the liquid-vapor binodal as well as the CP exists. But it is known both from the computer simulations and theoretical models that the region of the liquid phase may collapse if the range of interaction is too short (see \cite{eos_seconvircrit_jcp2000} and reference therein). This happen if the triple point meets the critical one. In such situation there is no critical state because the liquid branch is not the equilibrium phase in this situation.

Obviously the sufficient condition for the liquid branch to disappear as a stable state is $T_c \le T_{tr}$, where $T_{tr}$ is the triple point temperature. As the value of $z$ decreases with the decreasing the range of interaction the situation when $T_c$ occurs below $T_{tr}$ becomes possible. Therefore for the potential with varying range of interaction, e.g. via the power of the attractive part,
there is the value $z^*$ below which the liquid branch of the binodal ceases to exist as the stable phase.

From the point of view of the global isomorphism with the LG this could be interpreted as the moment when the interaction range becomes smaller than the lattice spacing of the effective LG model.

To estimate the value of $z^*$ we use the standard result that for LJ fluid $T_{tr}\approx 0.75\,T^{(LJ)}_{c} \simeq T^{(vdW)}_{B}/4$. Thus  from $T_{tr} = T_{c}$ we get the estimate  $z^{*} \approx 1/4$. Thus from \eqref{cp_locus_potential} we get:
\begin{equation}\label{cp_disapear}
\f{d\,\ln \left(\,\Phi(n^{-1/d}_c)/T_Z\,\right)}{d\,\ln n_c/n_b} = 4
\end{equation}
Applying the condition \eqref{cp_disapear} to the Sutherland potential \eqref{sutherland_potetial} and the Yukawa potential:
\begin{equation}\label{yukawa_potential}
\Phi(r) =
  \begin{cases}
    \infty\,, & \text{if}\quad r< \sigma \\
    -\,\varepsilon\,\f{\sigma}{r} \exp\left(\,-\lambda (r/\sigma -1) \,\right)\,, & \text{if} \quad r\ge \sigma\,,
  \end{cases}
\end{equation}
we get that the liquid branch loses its stability if $s>9$ and $\lambda > 11$ correspondingly. The computer simulations give the values $s\gtrsim 5$ and $\lambda \gtrsim 6$ \cite{eos_yukawafluid_jcp1994,crit_hardspheres_pre2003}. Obviously, the disagreement is caused by the crudeness of the estimate for the triple point temperature which was taken as for standard LJ fluid. In fact the position of the triple point also depends on the potential. The computer data correspond to the estimate $z^* \simeq 1/3$. This issue will be the subject of the separate paper.

%

\end{document}